\documentclass[prl, preprintnumbers ,showpacs,amsmath,amssymb, superscriptaddress,twocolumn]{revtex4-1}
\usepackage{bm}
\usepackage{amsmath}
\usepackage{amssymb}
\usepackage{amsthm}
\usepackage{amsfonts}
\usepackage{enumerate}
\usepackage{latexsym}
\usepackage{times}
\usepackage{ulem}
\usepackage{graphicx}
\usepackage{color}
\usepackage{makeidx}
\usepackage{ifpdf}

\begin{document}

\title{Two-dimensional \textit{J}$_{\rm eff}$ = 1/2 antiferromagnetic insulator unraveled from interlayer exchange coupling in artificial perovskite iridate superlattices}

\author{Lin~Hao}
\email{lhao3@utk.edu}
\affiliation{Department of Physics and Astronomy, University of Tennessee, Knoxville, Tennessee 37996, USA}
\author{D.~Meyers}
\affiliation{Department of Condensed Matter Physics and Materials Science, Brookhaven National Laboratory, Upton, New York 11973, USA}
\author{Clayton~Frederick}
\affiliation{Department of Physics and Astronomy, University of Tennessee, Knoxville, Tennessee 37996, USA}
\author{Gilberto~Fabbris}
\affiliation{Department of Condensed Matter Physics and Materials Science, Brookhaven National Laboratory, Upton, New York 11973, USA}
\author{Junyi~Yang}
\affiliation{Department of Physics and Astronomy, University of Tennessee, Knoxville, Tennessee 37996, USA}
\author{Nathan~Traynor}
\affiliation{Department of Physics and Astronomy, University of Tennessee, Knoxville, Tennessee 37996, USA}
\author{Lukas~Horak}
\affiliation{Department of Condensed Matter Physics, Charles University, Ke Karlovu 3, Prague 12116, Czech Republic}
\author{Dominik~Kriegner}
\affiliation{Department of Condensed Matter Physics, Charles University, Ke Karlovu 3, Prague 12116, Czech Republic}
\affiliation{Institute of Physics, Academy of Sciences of the Czech Republic, v.v.i., Cukrovarnická 10, 16253 Praha 6, Czech Republic}
\author{Yongseong~Choi}
\affiliation{Advanced Photon Source, Argonne National Laboratory, Argonne, Illinois 60439, USA}
\author{Jong-Woo~Kim}
\affiliation{Advanced Photon Source, Argonne National Laboratory, Argonne, Illinois 60439, USA}
\author{Daniel~Haskel}
\affiliation{Advanced Photon Source, Argonne National Laboratory, Argonne, Illinois 60439, USA}
\author{Phil~J. Ryan}
\affiliation{Advanced Photon Source, Argonne National Laboratory, Argonne, Illinois 60439, USA}
\affiliation{School of Physical Sciences, Dublin City University, Dublin 9, Ireland}
\author{M.~P.~M. Dean} \email{mdean@bnl.gov}
\affiliation{Department of Condensed Matter Physics and Materials Science, Brookhaven National Laboratory, Upton, New York 11973, USA}
\author{Jian~Liu}\email{jianliu@utk.edu}
\affiliation{Department of Physics and Astronomy, University of Tennessee, Knoxville, Tennessee 37996, USA}

\begin{abstract}
   We report an experimental investigation of the two-dimensional ${J}_{\rm eff}$ = 1/2 antiferromagnetic Mott insulator by varying the interlayer exchange coupling in [(SrIrO$_{3}$)$_{1}$, (SrTiO$_{3}$)$_{m}$] (${m}$ = 1, 2 and 3) superlattices. Although all samples exhibited an insulating ground state with long-range magnetic order, temperature-dependent resistivity measurements showed a stronger insulating behavior in the ${m}$ = 2 and ${m}$ = 3 samples than the ${m}$ = 1 sample which displayed a clear kink at the magnetic transition. This difference indicates that the blocking effect of the excessive SrTiO$_{3}$ layer enhances the effective electron-electron correlation and strengthens the Mott phase. The significant reduction of the Neel temperature from 150 K for ${m}$ = 1 to 40 K for ${m}$ = 2 demonstrates that the long-range order stability in the former is boosted by a substantial interlayer exchange coupling. Resonant x-ray magnetic scattering revealed that the interlayer exchange coupling has a switchable sign, depending on the SrTiO$_{3}$ layer number ${m}$, for maintaining canting-induced weak ferromagnetism. The nearly unaltered transition temperature between the ${m}$ = 2 and the ${m}$ = 3 demonstrated that we have realized a two-dimensional antiferromagnet at finite temperatures with diminishing interlayer exchange coupling.
\end{abstract}

\maketitle

A two-dimensional (2D) lattice formed of IrO$_{6}$ octahedra is the epitome of some of the most outstanding and challenging problems in condensed matter physics, such as electronic correlation with strong spin-orbit coupling (SOC), quantum magnetism, metal-insulator transition, doped Mott insulator, and latent superconductivity \cite{mar1,mar2,mar3,mar4,mar5,mar6,mar7,mar8,mar9,mar10,mar11,mar12,mar13}. The key notion permeating these emergent phenomena is the 2D pseudospin-half Mott insulating state stabilized through SOC that entangles $t_{2g}$ orbitals with spins and leads to a filled ${J}_{\rm eff}$ = 3/2 quartet and half-filled ${J}_{\rm eff}$ = 1/2 doublet \cite{mar2,mar6,mar14}. Of special interest is the square lattice of corner-sharing IrO$_{6}$ octahedra where the ${J}_{\rm eff}$ = 1/2 moments order in a 2D Heisenberg antiferromagnet as found in Sr$_{2}$IrO$_{4}$ and Ba$_{2}$IrO$_{4}$ with a Neel temperature of 240 K \cite{mar15,mar16,mar17,mar18}. In addition to the usual isotropic Heisenberg exchange, SOC is believed to cause significant anisotropic exchange interactions, i.e., pseudodipolar and Dzyaloshinskii-Moriya interactions, in the effective spin Hamiltonian \cite{mar1,mar19}. The former accounts for the basal-plane anisotropy in both compounds \cite{mar19}, whereas the latter induces large canting of the ${J}_{\rm eff}$ = 1/2 moments under the requisite octahedral rotations in Sr$_{2}$IrO$_{4}$ \cite{mar1}. Understanding such a 2D antiferromagnet is of great importance for quantum magnetism and its connection with high-temperature superconductivity \cite{mar20,mar21,mar22}. In fact, it has been proposed that doping these materials could lead to superconductivity \cite{mar4,mar23}. Recent studies have indeed found spectroscopic signatures similar to doped cuprates, including a $d$-wave electronic gap \cite{mar9,mar24,mar25,mar26} and persistent magnetic correlations \cite{mar27,mar28}.

\begin{figure}\vspace{-0pt}
\includegraphics[width=8cm]{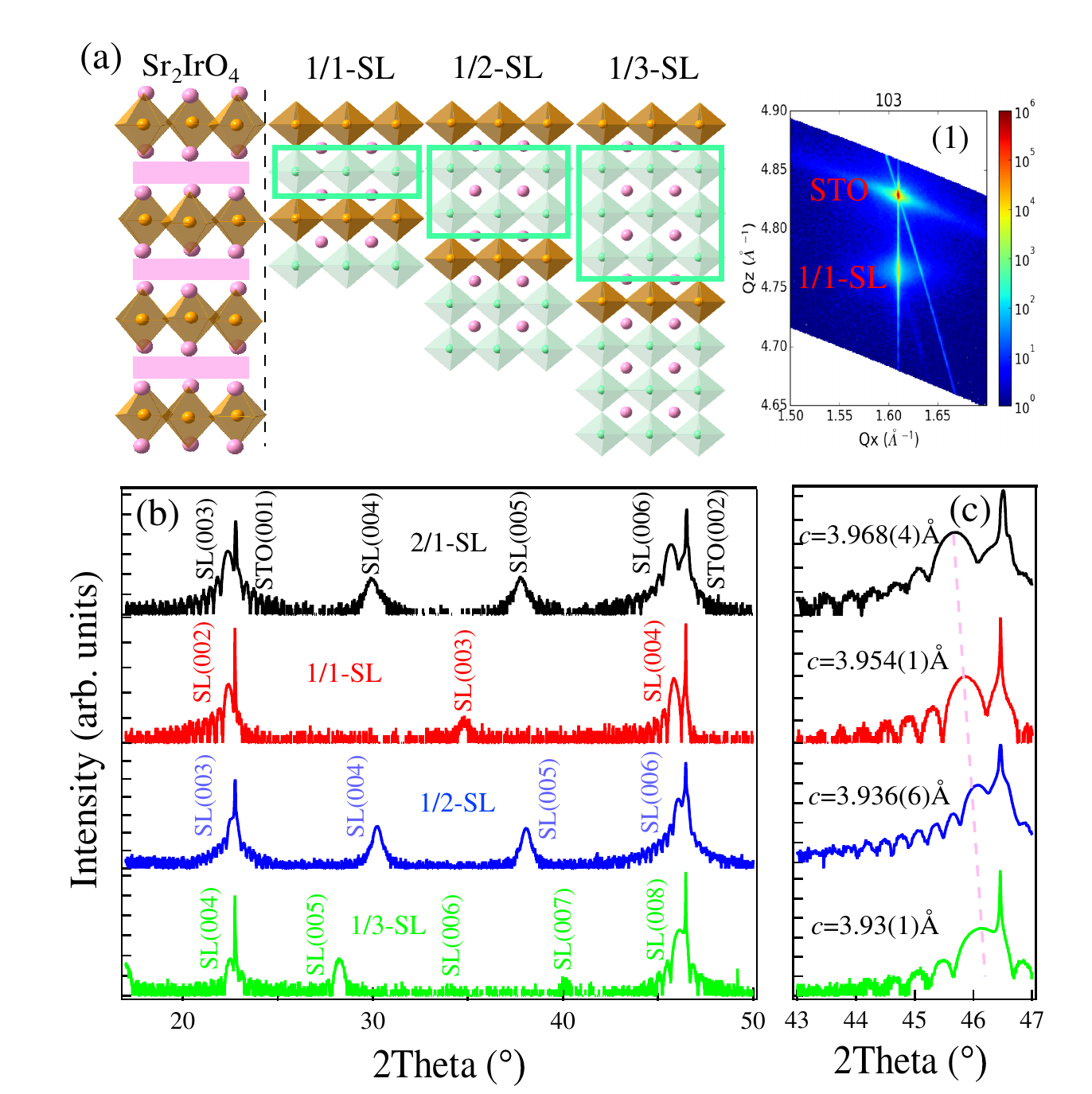}
\caption{\label{xrd} (a) Schematics of Sr$_{2}$IrO$_{4}$ and a series of SLs. The perovskite SIO layers are separated by a rock-salt SrO plane in the former and a STO block (green rectangle) in the latter. Ir$^{4+}$ (light orange) and Ti$^{4+}$ (light green) ions are surrounded by oxygen octahedra. (b) Room temperature $\theta-2\theta$ scans of the series of SLs. The corresponding SL peaks are also shown. (c) Magnified X-ray diffraction patterns around the STO (002) reflection. Inset (1) shows a representative reciprocal spacing mapping of a 1/1-SL around the (103) (pseudo-cubic lattice) film peak.}
\end{figure}

Previous investigations of square lattice ${J}_{\rm eff}$ = 1/2 materials were mostly on bulk Sr$_{2}$IrO$_{4}$ and Ba$_{2}$IrO$_{4}$, which are the ${n}$ = 1 end members of the Ruddlesden-Popper series, such as Sr$_{{n}+1}$Ir$_n$O$_{3n+1}$. The unit cell of Sr$_{n+1}$Ir${_{n}}$O$_{3{n}+1}$ can be considered as (SrIrO$_{3}$)$_{n}$\textbf{$\cdot$}SrO, which is composed of $n$ layers of perovskite SrIrO${_3}$ (SIO) sandwiched by rock-salt SrO monolayers that are considered to be electronically and magnetically inert \cite{mar12}. As the ${n}$ = $\infty$ end member, SIO displays exotic semi-metallic behavior due to the symmetry-protected Dirac line nodes \cite{mar29,mar30,mar31,mar32,mar33}. It was recently demonstrated that the layered lattice structure can be mimicked by replacing the SrO layers with SrTiO$_{3}$ (STO) layers, i.e., inserting a monolayer of STO in every ${n}$ layers of SIO during epitaxial growth \cite{mar34,mar35} (Fig.~\ref{xrd}(a)). As a wide band gap dielectric \cite{mar36}, the inserted layers of STO block the vertical charge hopping between SIO layers as its conduction band is $\sim$ 2 eV above the SIO ${J}_{\rm eff}$ = 1/2 band \cite{mar34}. The electric and magnetic ground states of the superlattices (SLs) [(SrIrO$_{3}$)$_{{n}}$, (SrTiO$_{3}$)$_{1}$] (${n}$ = 1, 2, 3, 4, and $\infty$) indeed exhibit an antiferromagnetic (AFM) insulator-to-paramagnetic metal crossover in analogy with the Sr$_{{n}+1}$Ir${_{n}}$O$_{3{n}+1}$ counterparts. Nevertheless, upon a closer examination, significant differences between the two series can be readily seen in their transition temperatures, magnetic structure, transport properties, optical conductivity, and their dimensional evolution \cite{mar34,mar37,mar38}. These differences indicate that the details of the blocking layer may play a critical role in the physical properties of the confined 2D SIO layers.

In this work, we investigated the interplay of the intralayer and interlayer couplings by varying the blocking layer thickness in [(SrIrO$_{3}$)$_{{n}}$, (SrTiO$_{3}$)$_{{m}}$] (${n}$ = 1 and 2 for ${m}$ = 1; ${m}$ = 1, 2 and 3 for ${n}$ = 1) SLs prepared through layer-by-layer epitaxial growth. The STO slab thickness in our SLs is highly tunable and can be precisely monitored during epitaxial deposition. This controllability of the blocking layer provides a unique benefit compared to studying the effects of the SrO layers in the Sr$_{{n}+1}$Ir$_n$O$_{3{n}+1}$ series, where the SrO layer is fixed by the equilibrium crystal growth. For simplicity, we hereafter use ${n}$/${m}$-SL to denote a SL with ${n}$(${m}$) successive SIO (STO) layers. For the 2/1- and 1/1-SLs, we observed AFM transitions and associated resistivity anomalies, consistent with the report in Ref. \cite{mar34}. Interestingly, while no such resistivity anomaly was found in the 1/2- and 1/3-SLs where the neighboring SIO layers are further separated, both samples still show a clear magnetic transition in magnetometry at lower temperatures. In addition, the transition temperatures of the 1/2- and 1/3-SLs were found to be almost the same, implying that the samples are approaching the 2D limit of the long-range magnetic order. The transition temperature was also verified by resonant x-ray magnetic scattering. The positions of the magnetic Bragg peaks reveal that the interlayer exchange coupling, although no longer significantly contributing to the Neel temperature, has a variable sign. This sign is entwined with the octahedral rotation relation of the adjacent SIO layers and remains effective in aligning the canted moments. This effect is crucial in arranging the local canting into the macroscopic weak ferromagnetism.

SLs with different stacking sequences were epitaxially grown by pulsed laser deposition on (001)-oriented STO substrates. A KrF excimer laser beam ($\lambda$ = 248 nm) with optimized fluency of 1.8 J/cm$^{2}$ was used to ablate stoichiometric targets. In situ reflection high energy electron diffraction (RHEED) patterns were monitored to control the film thickness in the range of 20 nm to 30 nm at the atomic level. During the deposition, the substrate temperature and O$_{2}$ pressure were set at 700 $^{o}$C and 0.1 mbar, respectively. The quality and structure of the SLs was checked by standard x-ray diffraction (XRD) using a Panalytical X$\rq$Pert MRD diffractometer. Magnetic measurements were performed on a Quantum Design superconducting quantum interference device magnetometer. The temperature-dependent electric resistivity was measured using a physical property measurement system. The magnetic x-ray scattering measurements were performed at Beamline 4IDD and 6IDD at the Advanced Photon Source of Argonne National Laboratory. The unit cell of ${a} \times {a} \times ({n}+{m}){a}$ (${a}$ is pseudo-cubic lattice parameter) was used to define the reciprocal space notation.

In Fig.~\ref{xrd}(b), only (0 0 ${L}$) reflections can be observed, indicating that all the SLs are epitaxially oriented. We also found characteristic SL peaks originating from the alternating stacking of the SIO and STO blocks along the ${c}$-axis, demonstrate that the SLs have been prepared as schematically shown in Fig.~\ref{xrd}(a). The well-defined Kiessig fringes [Fig.~\ref{xrd}(c)] demonstrate sharp substrate-film interfaces and flat surfaces. Additionally, there is a monotonic increase of film peak diffraction angle with decreasing ${n}$ or increasing ${m}$. The extracted ${c}$-axis (pseudocubic lattice) lattice parameters shows a linear dependence on the normalized layer composition ${n}/({n}+{m})$, corroborating the control of the layering growth. We then performed reciprocal space mapping (RSM) [see inset (1) in Fig.~\ref{xrd}] and found that all the films were coherently grown with the in-plane lattice parameter matched to the substrate. Further structural studies show that the all the samples have comparable and consistent qualities, including the same in-plane lattice parameter, mosaicity, and roughness.

\begin{figure}\vspace{-0pt}
\includegraphics[width=6.5cm]{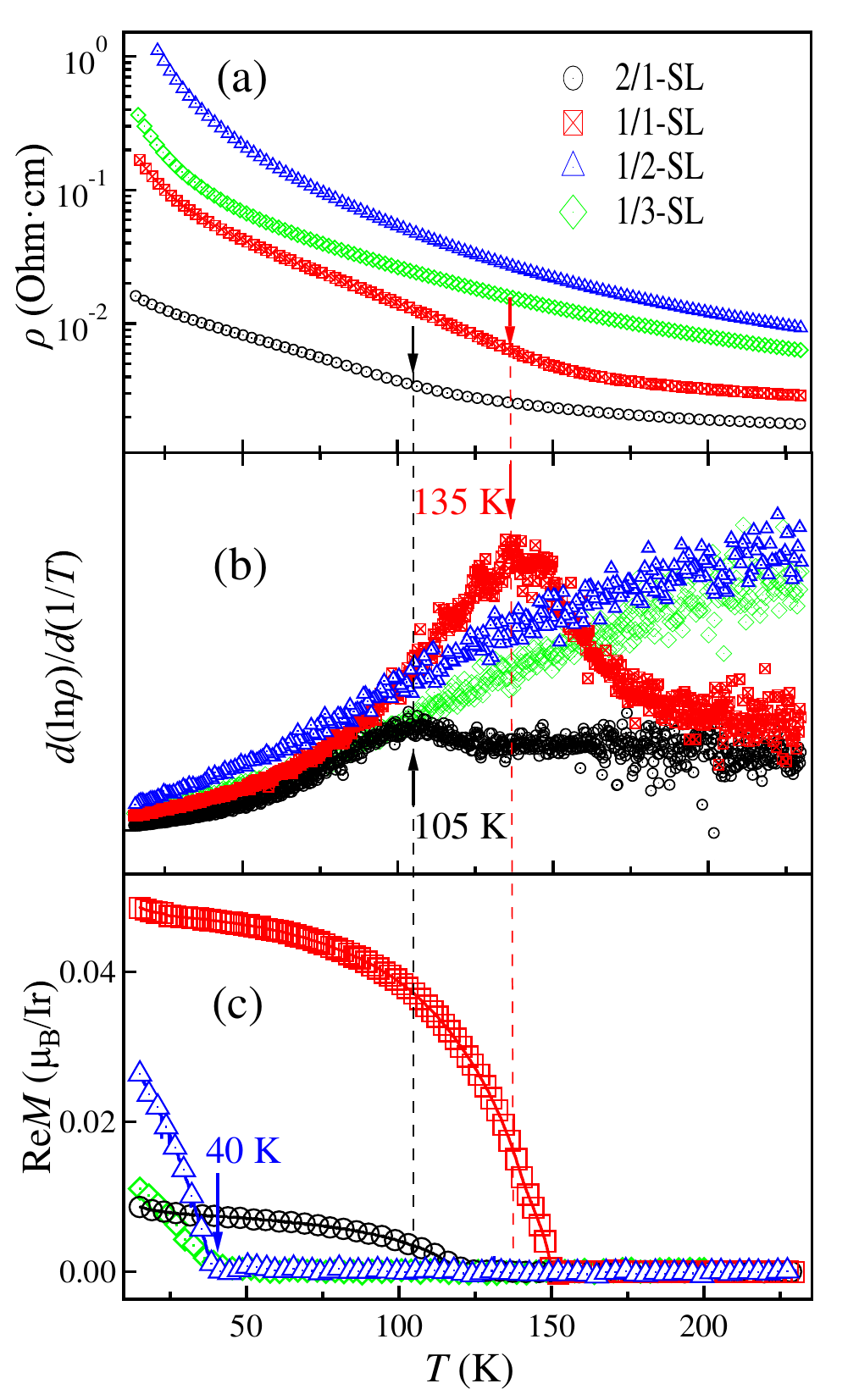}
\caption{\label{ppms} (a) Temperature dependent resistivity of the SLs. The arrows indicate the position of the resistivity anomaly in 2/1- and 1/1-SLs. (b) ${d}(ln\rho)/d(1/{T})$ shown as a function of ${T}$. (c) The remnant magnetization plotted against temperature. Before measurements under the zero-field condition, all the samples were cooled from room temperature to 2 K with application of a 5 kOe in-plane magnetic field. The dashed lines are guides to the eye.}
\end{figure}

Resistivity and magnetization measurements as a function of temperature afforded rich information on the emergent behavior the blocking layer affects. The temperature-dependence of resistivity $\rho$ is presented in Fig.~\ref{ppms}(a), in which insulating behavior can be established for all SLs. However, while resistivity anomalies at 105 K for 2/1-SL and 135 K for 1/1-SL were clearly observed, the $\rho-{T}$ plots of 1/2- and 1/3-SLs are smooth without any discernible kink. This can be further exhibited by investigating the relation between $d(ln\rho)/d(1/{T})$ and ${T}$ as shown in Fig.~\ref{ppms}(b). Again, no kink or peak was observed for the 1/2- or 1/3-SLs, while $\lambda$-like cusps can be easily seen for the 2/1- and 1/1-samples, characteristic of a phase transition. The observed kinks for ${m}$ = 1 are consistent with previous report \cite{mar34} and associated with the canted AFM transition. Figure~\ref{ppms}(c) shows the temperature dependent remnant magnetization (Re${M}$) of the 1/1- and 2/1-SLs which onsets at ${T}$ = 150 K and 120 K, respectively, and sharply increases close to the temperatures of their resistivity anomaly peaks. A recent optical conductivity study \cite{mar38} reported a larger bandwidth and a small charge gap in the 2/1- and 1/1-SLs compared with the Sr$_{{n}+1}$Ir$_n$O$_{3{n}+1}$ counterparts. This could allow the thermally activated carriers and their transport to be more easily subject to lower-energy magnetic excitations and reflected in the resistivity kinks upon long-range ordering. Since the reduced gap was attributed to additional effective Ir-Ir hopping across the thin STO blocking layer \cite{mar38}, one can expect that, when ${m}$ $>$ 1, the interlayer tunneling will be suppressed, reinforcing the effective electron-electron correlation. This picture is indeed consistent with the enhanced insulating behavior in the 1/2- and 1/3-SLs compared to the 1/1-SL, as seen from their larger logarithmic derivatives at high temperature well above the transitions [Fig.~\ref{ppms}(b)]

Driving the system to the 2D limit by suppressing the interlayer hopping also has a profound effect on the long-range magnetic order. We found that diminishing the interlayer exchange coupling significantly suppresses the AFM ordering temperature that, however, does not cease but is held with a finite value. Specifically, the temperature dependent Re${M}$ (Fig.~\ref{ppms}(c)) of the 1/2- as well as the 1/3-SLs onsets at about 40 K, indicating that the magnetic ordering survives at appreciable temperatures upon separating the SIO layers. The significant reduction of the transition temperature compared to that of 1/1 SL points to the nontrivial role of the interlayer exchange coupling in support of the quasi-2D long-range ordering in the 1/1-SL. This is well in line with the hindered electronic hopping and exchange coupling along the ${c}$-axis due to the insertion of additional STO blocking layers. The strong decrease of interlayer coupling in the 1/2-SL reveals its rapid decay with an increasing separation between the 2D magnetic layers, which is generally expected as the interlayer superexchange should decrease exponentially with the thickness of the blocking layer \cite{mar42}. Additionally, the very similar onset temperatures of the 1/3-SL and 1/2-SL suggest that the residual interlayer exchange coupling is sufficiently small and plays a secondary role in determining the transition temperature for ${m}$  $\geq$ 2. In other words, the long-range magnetic order in the 1/2- and 1/3-SLs is sustained by the easy-plane anisotropy due to the intralayer anisotropic exchange coupling.

This behavior also represents an exceptional macroscopic manifestation of magnetic anisotropy in stabilizing the spin-half 2D antiferromagnet at finite temperatures within the IrO$_{2}$ plane \cite{mar43}. It has been established that the dominant term in the magnetic Hamiltonian in Sr$_{2}$IrO$_{4}$ is an isotropic 2D Heisenberg coupling between ${J}_{\rm eff}$ = 1/2 moments \cite{mar4,mar5,mar44}. However, according to the Mermin-Wagner theorem \cite{mar45}, the long-range ordering of such an AFM state is unstable against thermal fluctuation at arbitrarily small finite temperatures. One possible route for the ordering to survive at finite temperature is by introducing anisotropy to the moments through crystal field distortion for example. We do not expect significant contribution from this route here due to symmetry protection of the Kramer’s doublet \cite{mar1}. In any case, the IrO$_6$ octahedron is elongated by about 4\% in Sr2IrO4 \cite{mar16} but rather uniform in perovskite SIO \cite{mar31}; thus, the weak orbital anisotropy in Sr$_2$IrO$_4$, would likely be weaker in SLs. Another route is through is through the easy-plane anisotropy by anisotropic exchange, which has been proposed to be nonnegligible in a 2D square lattice of IrO6 octahedra \cite{mar1,mar19}. Generally speaking, the transition temperature of such an anisotropic quantum antiferromagnet is set by and increasing with the easy-plane anisotropy characterized by the ratio between the anisotropic exchange and the Heisenberg
exchange \cite{mar46,mar47}. It has also been shown that the anisotropic exchange in iridates is typically stronger than cuprates due to the J ${J}_{\rm eff}$ = 1/2 spin-orbit wave function \cite{mar1,mar5,mar6,mar19}. A 40 K onset temperature of the ordering corresponds to an energy scale of order 3 meV, which implies that our SLs may have a larger anisotropy than Sr$_2$IrO$_4$, which has a small spin gap about 1 meV \cite{mar48,mar49}. We note that the SLs studied here have substantially different bond lengths and bond angles than Sr$_2$IrO$_4$ \cite{mar34,mar50}, and such a difference is not unexpected, especially in view of the large variations of the anisotropy in different iridates \cite{mar51,mar52}. The present observation is also reminiscent of the destruction of long-rang ordering in a single layer of La$_2$CuO$_4$ with the absence of interlayer coupling \cite{mar42}, despite the fact that the intralayer Heisenberg exchange is the dominant magnetic interaction in both systems.

\begin{figure}\vspace{-0pt}
\includegraphics[width=8.5cm]{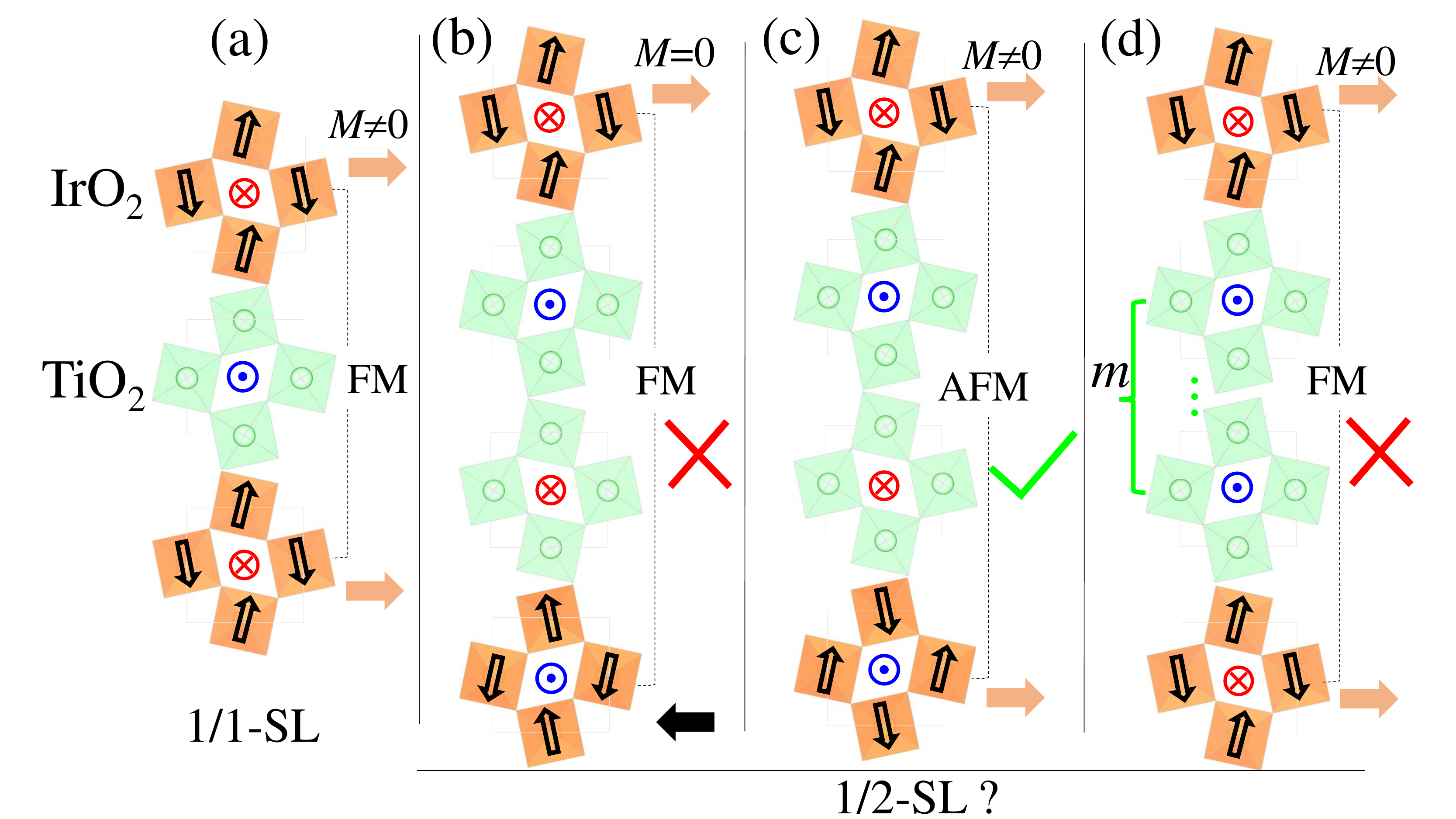}
\caption{\label{diagram} (a) The stacking pattern of the magnetic structure and octahedral rotation pattern of a 1/1-SL \cite{mar34}. (b), (c) and (d) Three possible magnetic structures of a 1/2-SL. TiO$_{6}$ octahedra within the STO block rotate out-of-phase in (b) and (c), while in-phase in (d). The magnetic structure in (c) with an AFM interlayer exchange coupling was verified by x-ray resonant magnetic scattering to be the correct one (Fig.~\ref{scattering}). The open and solid arrows denote ${J}_{\rm eff}$ = 1/2 moments and canted moments in each IrO$_{2}$ plane, respectively.}
\end{figure}

Although the former picture may illustrate the 2D AFM order within individual layers, one must also explain the observation of a nonzero net moment. Specifically, the net magnetizations in both Sr$_{2}$IrO$_{4}$ and the 1/1-SL originate from canting of the AFM moments within the plane \cite{mar2,mar34,mar53,mar54}, which is a consequence of the strong SOC that locks the AFM moments to the antiferrodistortive octahedral rotation \cite{mar1}. In addition, the canted moments must have a parallel interlayer alignment to avoid cancelling each other. In the ${n}$/1-SL, this parallel alignment was attributed to a combination of the ferromagnetic (FM) Ir-Ir interlayer exchange coupling and the out-of-phase rotation of neighboring octahedra along the ${c}$-axis (Fig.~\ref{diagram}(a)) \cite{mar34}, i.e., a ${c}^{-}$ rotation in Glazer notation \cite{mar55}. Since the overall behavior of our 1/1- and 2/1-SLs is almost identical to that in Ref. \cite{mar34}, we expect the same mechanism to be in play. It, however, breaks down when ${m}$ is increased. While the ${c}^{-}$ rotation ensures the adjacent SIO layers (intervened by STO layers) are in-phase when ${m}$ = 1 and 3 (Fig.~\ref{diagram}(a)), the adjacent SIO layers become out-of-phase when ${m}$ = 2. Assuming the FM Ir-Ir interlayer coupling persists in all SLs, the canting would follow the rotation phase and the canted moments of different layers would cancel each other in the 1/2-SL (Fig.~\ref{diagram}(b)), which is opposite to the observed net moment. To reconcile this, one may instead speculate an AFM Ir-Ir interlayer exchange coupling for ${m}$ = 2, which combined with the out-of-phase rotation would render a parallel alignment of the canted moments (Fig.~\ref{diagram}(c)). Another possible scenario is that the TiO$_{6}$ octahedral rotation is in-phase along the $c$-axis within the STO block but out-of-phase with the IrO$_{6}$ octahedra at the interfaces, as shown in Fig.~\ref{diagram}(a). In this case, the IrO$_{6}$ octahedral rotation of adjacent SIO layers is always in-phase regardless of the STO layer number ${m}$, and a FM Ir-Ir interlayer exchange coupling is valid.

\begin{figure}\vspace{-0pt}
\includegraphics[width=8.5cm]{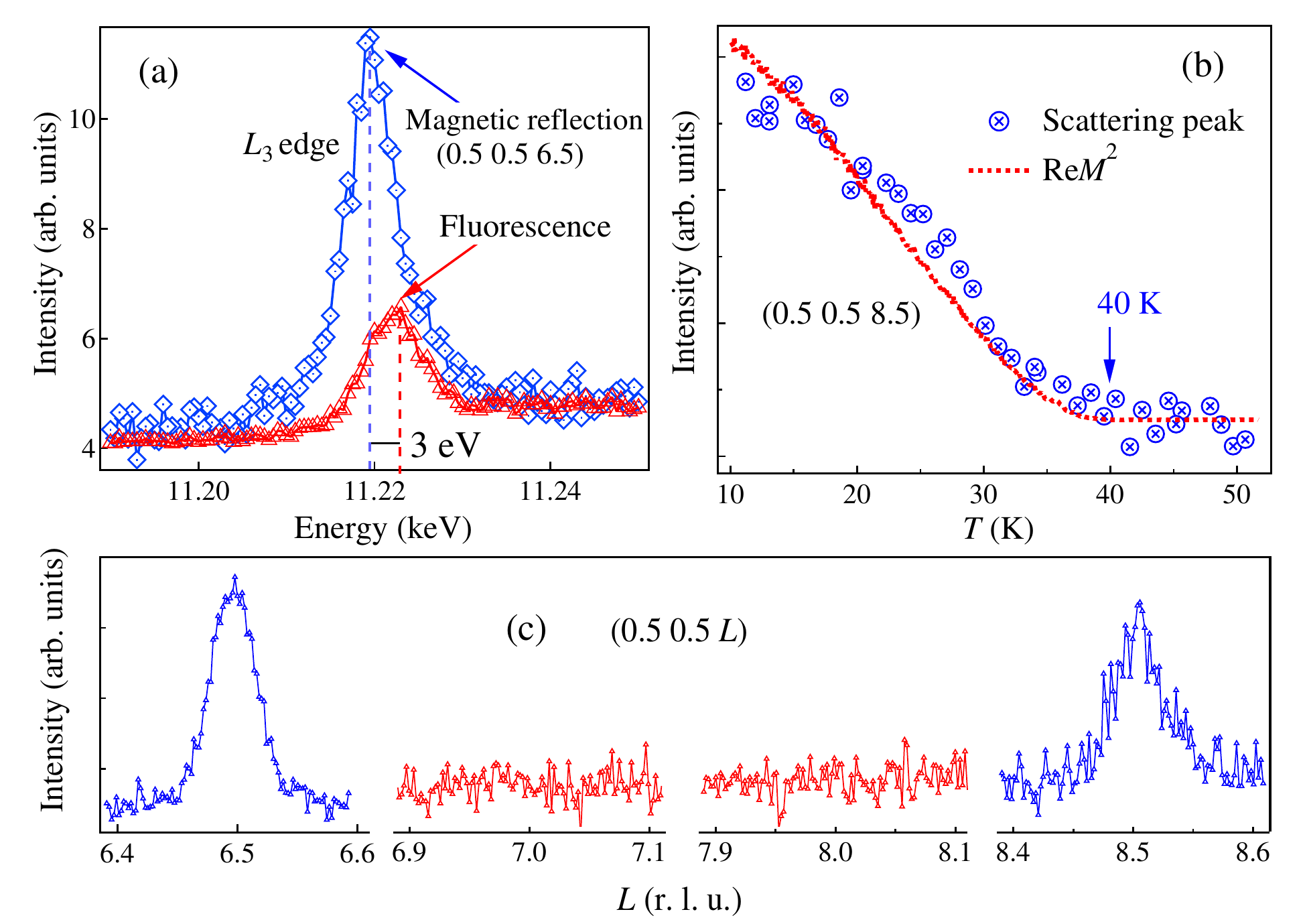}
\caption{\label{scattering} (a) Energy scan at the (0.5 0.5 6.5) magnetic reflection near the Ir ${L}_{3}$ edge at 10 K. The measured fluorescence is also shown for comparison. (b) Temperature dependent intensity of the (0.5 0.5 8.5) peak. It well reproduces the square of Re${M}$ as indicated by the dashed curve. (c) ${L}$-scans across the (0.5 0.5 6.5), (0.5 0.5 7.0), (0.5 0.5 8.0), and (0.5 0.5 8.5) magnetic reflections}
\end{figure}

To distinguish the latter two scenarios, a direct measurement of the interlayer exchange coupling is necessary since magnetometry is only sensitive to the net magnetic moment. Based on this consideration, we performed x-ray magnetic scattering measurement at the Ir ${L}_{3}$-edge on the 1/2-SL, in which the interlayer exchange coupling would be opposite between the two scenarios. An ${L}$-scan performed at the (0.5 0.5 ${L}$) magnetic Bragg reflection revealed that the nearest-neighboring ${J}_{\rm eff}$ = 1/2 moments within each IrO$_{2}$ plane are antiferromagnetically coupled, confirming the 2D AFM ground state. The energy profile (Fig.~\ref{scattering}(a)) of the peak intensity at 10 K also shows a typical lineshape with the maximum about 3 eV below the absorption peak represented by the fluorescence, which is similar to Sr$_{2}$IrO$_{4}$ \cite{mar2} and the 1/1-SL \cite{mar34}. Moreover, Fig.~\ref{scattering}(b) presents the temperature dependence of a magnetic Bragg peak, which increases at about 40 K, consistent with the magnetometry measurement and confirming the canting origin of the net moment. Most interestingly, we only observed magnetic Bragg peaks at ${L}$ = ${l}$+1/2, where ${l}$ is an integer (Fig.~\ref{scattering}(c)), indicative of a doubling of the magnetic unit cell along the ${c}$-axis with respect to the SL unit cell. This unambiguously shows that the Ir-Ir interlayer alignment is AFM, and confirms that the magnetic structure of the 1/2-SL should be as shown in Fig.~\ref{diagram}(c) (indicated with a check mark). For comparison, the FM interlayer exchange coupling in 1/1-SL was characterized by magnetic Bragg peaks only observed at ${L}$ = ${l}$ \cite{mar34}. This behavior indicates that interlayer exchange coupling can change sign depending on the phase relation of the SIO layers' rotation patterns and remains significant, although its magnitude no longer contributes to stabilizing the Neel temperature. In addition, the extracted interlayer coherence length of about 6 SL unit cells from the half width at half maximum ($\sim$0.026 r. l. u.) of the (0.5 0.5 6.5) peak is almost half of that for 1/1-SL (11 SL unit cells) \cite{mar34}. Such an enhanced fluctuation could be the reason for the reduced magnetization of the 1/2-SL from that of the 1/1-SL and Sr$_{2}$IrO$_{4}$.

In conclusion, we have tailored the spin-orbital physics in [(SrIrO$_{3}$)${_{n}}$, (SrTiO$_{3}$)${_{m}}$] SLs through atomic control of the SIO as well as the STO block thickness. By inserting one additional STO layer into the 1/1-SL, the effective electron-electron correlation is enhanced, while the interlayer exchange coupling is reduced, lowering the AFM transition temperature of the 1/2-SL as compared to the 1/1-SL. Upon further insertion of STO layers for the 1/3-SL, the ordering temperature remains at a similar value, suggesting that SOC-driven anisotropic intralayer exchange coupling is the driving force for their long-range 2D magnetic ordering. Combining these results with the x-ray magnetic scattering measurement, we found that the declining interlayer exchange coupling still plays a role in the net magnetization and its sign can be artificially modulated by varying the layer number in the STO block.

\section{Acknowledgments}
\begin{acknowledgments}
The authors acknowledge experimental assistance from H. D. Zhou, C. Rouleau, Z. Gai, J. K. Keum and H. Suriya. The authors would like to thank E. Dagotto, C. Batista, A. Eguiluz, and H. Xu for fruitful discussions. J.L. acknowledges the support by the Science Alliance Joint Directed Research $\&$ Development Program and the Transdisciplinary Academy Program at the University of Tennessee. J.L. also acknowledges the support by the DOD-DARPA under Grant No. HR0011-16-1-0005. M.P.M.D. is supported by the U.S. Department of Energy, Office of Basic Energy Sciences, Early Career Award Program under Award Number 1047478. Work at Brookhaven National Laboratory was supported by the U.S. Department of Energy, Office of Science, Office of Basic Energy Sciences, under Contract No. DESC00112704. D. K. and L. H. acknowledge support from the ERDF (Project No. CZ.02.1.01$/$0.0$/$0.0$/$15$\_$003$/$0000485) and the Grant Agency of the Czech Republic Grant No. (14-37427G). A portion of fabrication and characterization was conducted at the Center for Nanophase Materials Sciences, which is a DOE Office of Science User Facility. Use of the Advanced Photon Source, an Office of Science User Facility operated for the U. S. DOE, OS by Argonne National Laboratory, was supported by the U. S. DOE under Contract No. DE-AC02-06CH11357.
\end{acknowledgments}


\end{document}